\documentclass[a4paper,12pt, epsfig]{article}
\usepackage{epsfig}
\usepackage{amssymb}
\usepackage{amsfonts}

\newskip\humongous \humongous=0pt plus 1000pt minus 1000pt

\newif\ifdtup

\catcode`\@=11

\@addtoreset{equation}{section}
\def\theequation{\thesection.\arabic{equation}}

\def\@normalsize{\@setsize\normalsize{15pt}\xiipt\@xiipt
\abovedisplayskip 14pt plus3pt minus3pt%
\belowdisplayskip \abovedisplayskip
\abovedisplayshortskip \z@ plus3pt%
\belowdisplayshortskip 7pt plus3.5pt minus0pt}

\def\small{\@setsize\small{13.6pt}\xipt\@xipt
\abovedisplayskip 13pt plus3pt minus3pt%
\belowdisplayskip \abovedisplayskip
\abovedisplayshortskip \z@ plus3pt%
\belowdisplayshortskip 7pt plus3.5pt minus0pt
\def\@listi{\parsep 4.5pt plus 2pt minus 1pt
     \itemsep \parsep
     \topsep 9pt plus 3pt minus 3pt}}

\relax

\catcode`@=12

\catcode`\@=11

\def\section{\@startsection{section}{1}{\z@}{3.5ex plus 1ex minus
   .2ex}{2.3ex plus .2ex}{\large\bf}}

\def\thesection{\arabic{section}}
\def\thesubsection{\arabic{section}.\arabic{subsection}}

\def\appendix{\setcounter{section}{0}
 \def\thesection{Appendix \Alph{section}}
 \def\thesubsection{\Alph{section}.\arabic{subsection}}
 \def\theequation{\Alph{section}.\arabic{equation}}}

\def\SymBoxes#1#2#3#4{\newdimen\un@t \un@t#3%
\raisebox{#1}{\rule{#2\un@t}{#4}\hskip-#2\un@t
\@tempdimb\un@t \advance\@tempdimb by-#4\@tempcntb#2\relax%
\@whilenum{\@tempcntb>0}\do{
\rule{#4}{\un@t}\hskip\@tempdimb \advance\@tempcntb by\m@ne}%
\hskip-#2\un@t \rule[\un@t]{#2\un@t}{#4}%
\rule[\un@t]{#4}{#4}\hskip-#4
\rule{#4}{\un@t}}\hskip-#4}                

\begin{document}

\newcommand{\beq}{\begin{equation}}
\newcommand{\eeq}{\end{equation}}
\newcommand{\bea}{\begin{eqnarray}}
\newcommand{\eea}{\end{eqnarray}}
\newcommand{\beas}{\begin{eqnarray*}}
\newcommand{\eeas}{\end{eqnarray*}}
\newcommand{\defi}{\stackrel{\rm def}{=}}
\newcommand{\non}{\nonumber}
\newcommand{\bquo}{\begin{quote}}
\newcommand{\enqu}{\end{quote}}
\renewcommand{\(}{\begin{equation}}
\renewcommand{\)}{\end{equation}}
\def\IZ{{\mathbb Z}}
\def\IR{{\mathbb R}}
\def\IC{{\mathbb C}}
\def\IQ{{\mathbb Q}}
\def\Rhat{{\hat R}}
\def\Chat{{\hat C}}

\def \eqn#1#2{\begin{equation}#2\label{#1}\end{equation}}
\def\de{\partial}
\def\Tr{ \hbox{\rm Tr}}
\def\H{ \hbox{\rm H}}
\def\HE{ \hbox{$\rm H^{even}$}}
\def\HO{ \hbox{$\rm H^{odd}$}}
\def\K{ \hbox{\rm K}}
\def\Im{ \hbox{\rm Im}}
\def\Ker{ \hbox{\rm Ker}}
\def\const{\hbox {\rm const.}}
\def\o{\over}
\def\im{\hbox{\rm Im}}
\def\re{\hbox{\rm Re}}
\def\bra{\langle}\def\ket{\rangle}
\def\Arg{\hbox {\rm Arg}}
\def\Re{\hbox {\rm Re}}
\def\Im{\hbox {\rm Im}}
\def\exo{\hbox {\rm exp}}
\def\diag{\hbox{\rm diag}}
\def\longvert{{\rule[-2mm]{0.1mm}{7mm}}\,}
\def\a{\alpha}
\def\dag{{}^{\dagger}}
\def\tq{{\widetilde q}}
\def\p{{}^{\prime}}
\def\W{W}
\def\N{{\cal N}}
\def\hsp{,\hspace{.7cm}}
\newcommand{\C}{\ensuremath{\mathbb C}}
\newcommand{\Z}{\ensuremath{\mathbb Z}}
\newcommand{\R}{\ensuremath{\mathbb R}}
\newcommand{\rp}{\ensuremath{\mathbb {RP}}}
\newcommand{\cp}{\ensuremath{\mathbb {CP}}}
\newcommand{\vac}{\ensuremath{|0\rangle}}
\newcommand{\vact}{\ensuremath{|00\rangle}                    }
\newcommand{\oc}{\ensuremath{\overline{c}}}
\begin{titlepage}
\begin{flushright}
\end{flushright}
\bigskip
\def\thefootnote{\fnsymbol{footnote}}

\begin{center}
{\Large {\bf
Can Quantum de Sitter Space Have\\  
\vspace{0.3em}
Finite Entropy?}} \\ 
\end{center}

\bigskip
\begin{center}
{\large  Chethan 
KRISHNAN$^{1}$\footnote{\texttt{Chethan.Krishnan@ulb.ac.be}} 
and Edoardo DI NAPOLI$^{2}$
\footnote{\texttt{edodin@physics.unc.edu}}}\\
\end{center}

\renewcommand{\thefootnote}{\arabic{footnote}}

\begin{center}
\vspace{1em}
{\em  $^{1}${ International Solvay Institutes,\\
Physique Th\'eorique et Math\'ematique,\\
ULB C.P. 231, Universit\'e Libre
de Bruxelles, \\ B-1050, Bruxelles, Belgium\\}}
\vspace{1em}
{\em $^{2}${ Department of Physics and Astronomy \\
CB\# 3255 Phillips Hall \\
University of North Carolina \\
Chapel Hill, NC 27599-3255, USA \\}}

\end{center}

\noindent
\begin{center} {\bf Abstract} \end{center}
 If one tries to view de Sitter as a true (as opposed to a meta-stable) 
vacuum, there is a tension between the finiteness of its entropy and the 
infinite-dimensionality of its Hilbert space. We invetsigate the viability 
of one proposal to reconcile this tension using $q$-deformation. After 
defining a differential geometry on the quantum de Sitter space, we try to 
constrain the value of the deformation parameter by imposing the condition 
that in the undeformed limit, we want the real form of the (inherently 
complex) quantum group to reduce to the usual SO(4,1) of de Sitter. We 
find that this forces $q$ to be a real number. Since it is known that 
quantum groups have finite-dimensional representations only for $q=$ root 
of unity, this suggests that standard $q$-deformations cannot give rise to 
finite dimensional Hilbert spaces, ruling out finite entropy for 
$q$-deformed de Sitter.

\vspace{1.6 cm}

\vfill

\end{titlepage}
\bigskip

\hfill{}
\bigskip

\tableofcontents

\setcounter{footnote}{0}



\section{\bf Introduction}

Observations indicate that our Universe is currently in a regime of 
accelerated expansion, and that we might be living in an asymptotically 
de Sitter spacetime \cite{Willy, Willy2}. From the perspective of a 
co-moving 
observer, de Sitter 
spacetime has a cosmological horizon and an associated finite entropy 
\cite{Willy, Bousso}. 
Finiteness of entropy suggests finite dimensionality of the Hilbert space, 
but since the isometry group of de Sitter is non-compact and therefore has 
no 
finite dimensional unitary representations, we immediately have a problem 
in our hands \cite{Banks, Willy3}.

One idea that has been proposed as a way out of this quandary is to look for 
$q$-deformations \cite{Pouliot, Lowe1, Lowe2, KN} of the isometry group 
which might admit finite 
dimensional unitary representations. When the deformation parameter(s) are 
taken to $\rightarrow 1$, we recover the classical group. It is known that 
the standard $q$-deformations of (complexified and therefore non-compact) 
classical groups have 
finite dimensional unitary representations when $q$ is a root of 
unity \cite{dobrev1, dobrev2, dobrev3, Stein1, Stein2, Stein3, Klimyk}. 

In this paper, we first look at how the geometric structure of the 
underlying de Sitter space is modified when its symmetry group is 
deformed. It turns out that a deformed 
symmetry group necessitates a deformed differential calculus on the 
underlying space in order for the differential structure to be covariant 
with respect to the new $q$-symmetry. In section 2, we will make use of the 
work by Zumino et al. \cite{Zumino1, Zumino2} to explicitly write down the 
differential calculus 
on the quantum Euclidean space underlying the deformed
$SO(5;{\IC})$. Quantum groups are defined through their 
actions\footnote{To be 
precise we should say co-actions, but we are using the word loosely.} on 
complex vector spaces, and so we need to start with $SO(5;{\IC})$ before 
restricting to an appropriate real form to obtain $SO(4,1)$.

To obtain this real form, we need to choose a ``$*$-structure" 
(conjugation) on the algebra \cite{FRT, Aschieri}. We do this 
in 
section 3 and get 
$SO(4,1)_q$. The definition of 
conjugation that is necessary for imposing the reality 
condition on the group elements, induces a conjugation on the
underlying quantum space as well. Using this we can choose our 
co-ordinates to be real, and by imposing an $SO(4,1)_q$-covariant 
constraint on the quantum space we get a definition of quantum deSitter 
space. This is analogous to the imposition of 
$-(X^0)^2+(X^i)^2=1$ on a five-dimensional Minkowski space (thought of as 
a normed real vector space) to get the classical de Sitter space. 

The interesting thing is that the allowed real form of $SO(5;{\IC})_q$ 
which 
gives rise to $SO(4,1)$ in the $q \rightarrow 1$ limit is constrained by 
the condition that $q$ be real. But the representation theory of 
standard quantum groups allows finite-dimensional representations only 
when the deformation parameter is a root of unity \cite{Klimyk, Rosso}. 
This 
suggests that 
to get finite dimensional Hilbert spaces that could possibly be useful 
for de Sitter physics, we might need to look for non-standard 
deformations. 
Or it might be an indication that quantum mechanics in de Sitter space is 
too pathological to make sense even after $q$-deformation. 

Finiteness of de Sitter Hilbert space has also been discussed in 
\cite{Parikh:2004ux, Parikh:2004wh}, and $q$-deformation in the context of 
AdS/CFT has been considered in \cite{Jevicki:2000ty, Corley:2002ny}.

\section{\bf Differential Calculus on the Quantum Euclidean Space}

Following \cite{Zumino1}, we will consider deformations of the 
differential 
structure of the underlying space (with co-ordinates $x^k$, 
$k=1,2,...N$) by introducing matrices 
$B$, $C$ and $F$ (built of numerical coefficients) such that
\begin{eqnarray}\label{deformedcalc}
B^{kl}_{mn}x^mx^n&=&0, \\
\label{d2}\partial_lx^k&=&\delta_l^k+C^{km}_{ln}x^n\partial_m, \\
\label{d3}\partial_n\partial_m F^{mn}_{kl}&=&0.
\end{eqnarray}
In the limit when there is no deformation, these matrices should tend to 
the limits
\begin{eqnarray}\label{limit}
B^{kl}_{mn} &\rightarrow& (\delta^k_m\delta^l_n-\delta^l_m\delta^k_n), \\
C^{lk}_{nm} &\rightarrow& \delta^k_n\delta^m_l, \\
F^{mn}_{kl} &\rightarrow& (\delta^n_k\delta^m_l-\delta^n_l\delta^m_k),
\end{eqnarray}
so that we have the usual algebra of coordinates and their derivatives. We 
could also deform the commutation relations for 
1-forms and exterior differentials, but these follow straightforwardly 
from the matrices $B$, $C$ and $F$ upon imposing natural properties like 
Leibniz rule etc. So we will not concern ourselves with them here.

To construct a calculus on the space that is covariant under the 
co-action of a quantum group, we will use the $R$-matrix of the 
appropriate quantum group to define our matrices $B$, $C$ and $F$. To do 
this, we first look at the matrix\footnote{$\Rhat$ is also often called 
the $R$-matrix, but we will not do so to avoid confusion.} $\Rhat$, which 
is 
related to the $R$-matrix through $\Rhat^{ij}_{kl} \equiv R^{ji}_{kl}$. 
This $\Rhat$ satisfies the quantum Yang-Baxter equation by virtue of the 
fact that $R$ does:
\eqn{QYB}{\Rhat_{12}\Rhat_{23}\Rhat_{12}=\Rhat_{23}\Rhat_{12}\Rhat_{23}.}
It has a characteristic equation of the form:
\eqn{characteristic}{(\Rhat-\mu_1 I)(\Rhat-\mu_2 I)...(\Rhat-\mu_m I)=0.}  
It turns out that we can meet all the consistency requirements that 
$B$, $C$ and $F$ should satisfy in order for them to define a consistent 
deformation, if we set
\begin{eqnarray}
C&=&-\Rhat/\mu_{\alpha},\label{C} \\
B=F&=&\prod_{\beta (\neq \alpha)}(\Rhat-\mu_{\beta}I)\label{BF},
\end{eqnarray}
with some choice of the eigenvalue $\mu_{\alpha}$. With these definitions, 
the consistency conditions become 
automatic because $\Rhat$ satisfies the Yang-Baxter equation.

The 
$R$-matrix for $SO(2n+1)$ 
\cite{FRT} looks like
\begin{eqnarray}
\nonumber R&=&q\sum_{i\neq i^{\prime}}^{2n}E_{ii}\otimes
E_{ii}+q^{-1}\sum_{i\neq i^{\prime}}^{2n}E_{ii}\otimes 
E_{i^{\prime}i^{\prime}}
+E_{n+1,n+1}\otimes E_{n+1,n+1}+ \\
&+&\sum_{i \neq j, j^{\prime}}^{2n}E_{ii}\otimes E_{jj} + (q-q^{-1})\Big[ 
\sum_{i>j}^{2n}E_{ij}\otimes E_{ji} -
\sum_{i>j}^{2n}q^{\rho_i-\rho_j}E_{ij}\otimes E_{i^{\prime}j^{\prime}} 
\Big].
\end{eqnarray}
For $SO(5)$, $n=2$, $i$ and $j$ run from 1 to 5, and $E_{ij}$ is 
the $5 \times 5$ matrix with 1 
in the $(i,j)$-position and 0
everywhere else. The symbol $\otimes$ stands for tensoring of two 
matrices. We define $i^{\prime}=6-i$ and $j^{\prime}=6-j$. The deformation 
parameter is $q$.
Finally, 
$(\rho_1,\rho_2,...,\rho_{5})=(3/2,
1/2,0,-1/2,-3/2)$. 

The quantum group is defined in terms of 
matrices $T=(t_{ij})$ so that $RT_1T_2=T_2T_1R$ where $T_1=T\otimes I$ and 
$T_2=I \otimes T$. For deforming orthogonal groups we also 
need to specify a norm that is left invariant under the quantum group 
elements. This is done through the introduction of the matrix $\Chat$ 
(not to be confused with the $C$ introduced earlier) so that $T^t\Chat 
T=T\Chat T^t=\Chat$ where (for the specific case of $SO(5)$) 
\eqn{Chat}{\Chat=\left(\begin{array}{ccccc}
                 & & & &q^{-3/2}\\
                 & & &q^{-1/2}& \\
                 & &1& &        \\
                 &q^{1/2}& & &  \\
                q^{3/2}& & & &\end{array}\right).~}
We can use an $SO(5)_{q}$ invariant constraint of the form $
x^t\Chat x=$constant (where $x=\{x^i\}$) to define invariant 
subspaces of the Euclidean space. With appropriate reality conditions, 
this can give rise to different signatures in the classical limit.

Using the $R$-matrix defined above, we can define the $\Rhat$ matrix for 
our quantum 
orthogonal space and it has three distinct eigenvalues: 
$1/q^4, 
-1/q$ and $q$. 
By explicit 
computation using (\ref{C}, \ref{BF}) and 
(\ref{deformedcalc}, \ref{d2}, \ref{d3}) we find that the choice of 
$\mu_\alpha$ that gives rise to a non-degenerate deformation is
$\mu_{\alpha}=-1/q$. 
We write down this algebra explicitly in an appendix. 

This and the other computations done in this article
were implemented using the Mathematica package NCALGEBRA (version 
3.7) \cite{NC}.
	
\section{\bf Choice of Real Form}

So far we have worked with complexified groups and their deformations. 
But since we are interested in $SO(4,1)$ which is a specific real form 
of $SO(5;{\IC})$, we need to impose a reality condition on the $q$-group 
elements. For that, we need a definition of conjugation ($*$-structure). 
The $*$-structure on the quantum group will induce a conjugation 
on the underlying quantum space, and we want our co-ordinates to be real 
under this conjugation. In a basis where the co-ordinates and the quantum 
group elements are real, we can write down 
the metric $\Chat$. If the signature of that metric in the $q \rightarrow 
1$ limit is $\{-++++\}$, we have the real form that we are looking for. 
This is the program we will resort to, for writing down $SO(4,1)_q$. In 
this section, we will be working in the single-parameter context.

Using \cite{FRT, Aschieri}, we define a $*$-structure\footnote{A 
conjugation
$*$ on
a Hopf algebra $A$ is an algebra anti-automorphism, i.e., 
$*(\eta a.b)={\bar \eta} (*(b)).(*(a))$ for all $a,b \in A$ and $\eta
\in {\IC}$, that also happens to be a co-algebra automorphism,
$\Delta(*)=(* \otimes *)(\Delta)$,
$\epsilon(*)= \epsilon$ and an involution, $*^2=$identity.}
 by the relation 
\eqn{star}{T^{*}\equiv D\Chat^tT\Chat^tD^{-1}} where 
$$D=\left(\begin{array}{ccccc}
                 1& & & & \\
                  &1& & & \\
                  & &-1&& \\
                  & & &1& \\
                  & & & &1\end{array}\right)~.$$
The idea here is this: from FRT \cite{FRT}, we know of the conjugation 
$\star$, which 
is defined by $T^{\star}\equiv \Chat^tT\Chat^t$. Since it is known that 
$\star$ does not lead to the real form that we are looking for, we use the 
involution $D$, to create a new conjugation from $\star$. 
$D$ can be 
shown to respect all Hopf algebra structures: it is a Hopf algebra 
automorphism. Before we go further, it 
should also be mentioned that in 
order for the conjugation 
$\star$ to preserve the $RTT$ equations, the $R$-matrix should satisfy 
${\bar R}^{ij}_{kl}=R^{lk}_{ji}$ which works if $q \in {\IR}$.	

Using the fact that quantum groups co-act on the quantum space, we can 
induce a conjugation on the quantum space which turns out to be 
$x^*=\Chat^t D x$. As per our program, our next step would be 
to find a linear transformation
\begin{eqnarray}
x \rightarrow x'&=&Mx \\
T \rightarrow T'&=&MTM^{-1} 
\end{eqnarray}
such that $x', T'$ are real under their respective conjugations. Under 
such a transformation, the metric $\Chat$ must go to
\eqn{transC}{\Chat \rightarrow {\Chat}'=(M^{-1})^t\Chat M^{-1}.} 
One can check that an $M$ that can do this is, 
$$M=\frac{1}{\sqrt{2}}\left(\begin{array}{ccccc}
                            1&  &  &  &1\\
                             & 1&  &1 &\\
                             &  &i\sqrt{2}& &\\
                             & i&  &-i&  \\
                            i&  &  &  &-i \end{array}\right)~$$
and under it, the matrix $\Chat$ goes to
$$\Chat '=\left(\begin{array}{ccccc}
                      \frac{1}{2q^{3/2}}+\frac{q^{3/2}}{2}& 0 & 0& 0 
&\frac{i}{2q^{3/2}}-\frac{iq^{3/2}}{2}\\
                            0 & 
\frac{1}{2q^{1/2}}+\frac{q^{1/2}}{2}&0&\frac{i}{2q^{1/2}}-\frac{iq^{1/2}}{2} 
&0\\
                            0 & 0&-1&0 &0\\
                            0 & 
-\frac{i}{2q^{1/2}}+\frac{iq^{1/2}}{2}&0&\frac{1}{2q^{1/2}}+\frac{q^{1/2}}{2}&0  
\\
                            -\frac{i}{2q^{3/2}}+\frac{iq^{3/2}}{2}& 0 & 0 
&0&\frac{1}{2q^{3/2}}+\frac{q^{3/2}}{2} \end{array}\right)~.$$
It is clear that when $q \rightarrow 1$, we get the correct signature with 
one negative and four positive eigenvalues. So we finally have a 
complete definition of our quantum de Sitter space.

We note here that it is not obvious that all the real forms of the 
classical groups exist  
after $q$-deformation. Twietmeyer \cite{Twiet} and Aschieri 
\cite{Aschieri} have classified 
the possible real forms of $SO(2n+1)_q$, and they find that for real 
values of $q$ there are $2^n$ real forms (our de Sitter belongs to this 
category), and for $|q|=1$ there is only one real form, namely 
$SO(n,n+1)_q$. Since we expect to get finite dimensional representations 
of quantum groups only when $q$ is a root of unity\footnote{At other 
values 
of $q$, the representation theory of quantum groups is ``pretty much 
isomorphic" to the representation theory of classical groups.}, for 
$SO(5)$, these are allowed 
only for anti-de Sitter space \cite{Stein1, Stein2, Stein3}: with isometry 
group $SO(2,3)$. DeSitter symmetry group can occur only if we choose $q$ 
to be real, as we saw explicitly.
	
As already stressed, it might be possible to skirt this issue by working 
with more generic (multi-parametric or otherwise) deformations and 
corresponding $*$-structures. But we will not be pursuing those lines 
here. Maybe de Sitter should only be looked at as a resonance or 
a metastable 
state in some fundamental theory like string theory.
 
\section{Acknowledgments}

It is a pleasure to thank Willy Fischler for stimulating conversations and 
encouragement. We also wish to thank Jeff Olson, his help was 
crucial in cutting short the learning curve with Mathematica tricks. This 
material is based upon work supported by the National Science 
Foundation under Grant Nos. PHY-0071512 and PHY-0455649, and with grant 
support from the US Navy, Office of Naval Research, Grant Nos. 
N00014-03-1-0639 and N00014-04-1-0336, Quantum Optics
Initiative.

\section{\bf Appendix}

In this appendix we explicitly write down the form of the 
$SO(5)_{q}$-covariant differential calculus with  
$\mu_{\alpha}=-1/q$ as our chosen eigenvalue. The technology is well-known 
in the literature, we write down the explicit form here so we can 
make 
some comments. 

We first write down the commutators between the coordinates $x^i$ where 
$i$ goes from 1 to 5. 
\begin{eqnarray}
x^1x^2=\frac{1}{q} x^2x^1, \ x^2x^5=\frac{1}{q} x^5x^2 \\
x^1x^4= \frac{1}{q} x^4x^1, \ x^4x^5=\frac{1}{q} x^4x^5 \\
x^1x^3=qx^3x^1, \ x^2x^3=qx^3x^2 , \ x^3x^4=qx^4x^3, \ x^3x^5=qx^5x^3 \\
q(x^1x^5-x^5x^1)+(q-1)(x^2x^4+qx^4x^2)+q^{1/2}(q-1)x^3x^3=0 \\
q(q-1)(x^1x^5-x^5x^1)+(1-q+q^3)x^2x^4-q^2x^4x^2+q^{5/2}(q-1)x^3x^3=0 \\
\label{check} 
q^{3/2}(x^1x^5-x^5x^1)+q^{1/2}(q^2x^2x^4-x^4x^2)+(q^3-q^2+q-1)x^3x^3=0 \\
q^{3/2}(q-1)(x^1x^5-x^5x^1)-q^{3/2}x^2x^4+q^{1/2}(1-q^2+q^3)x^4x^2-(q-1)x^3x^3=0
\end{eqnarray}

Some of the relations above are redundant. Also, these relations should be 
thought of in conjunction with the condition that  $x^t \Chat 
x=\frac{3}{\Lambda}$ which (with the appropriate reality condition) is 
the embedding corresponding to de Sitter space. Here $\Lambda$ could be 
interpreted as the cosmological constant. One can rewrite the relations 
above by making use of this constraint and eliminating 
$x^3x^3$. For instance, 
after some algebra 
(\ref{check}) becomes
\begin{eqnarray}
\Big(\frac{1}{q^2}x^1x^5-q^2x^1x^5\Big)+\Big(\frac{1}{q}x^2x^4-qx^4x^2\Big)=\frac{\Lambda(1-q^4)}{3q^{1/2}(1+q^3)}.
\end{eqnarray}

It should be noted that these deformed commutators constructed a 'la 
Zumino, are the same as the ones written down by \cite{FRT} 
and \cite{Watamura}. 
Their prescription was to split 
the $R$-matrix into 
projection operators so that
\begin{eqnarray}
\Rhat\equiv q P_S-q^{-1}P_A+q^{-4}P_1,
\end{eqnarray}
and use that to define the deformations according to $P_A({\bf x}\otimes 
{\bf 
x})=0$. The projectors are onto the eigen-subspaces, so everything is 
consistent. 

The algebra of the partial derivatives (which is controlled by the 
matrix $F$) can be obtained from the co-ordinate algebra above 
if we substitute $x^i \rightarrow \partial_{x^{i'}}$, where $i'=6-i$.

To complete the definition of the deformed calculus, we need to spell out 
the 
algebra that deforms the commutators between the co-ordinates and 
derivatives. It turns out that they are, for $x^i$ and 
$\partial_{x^j}$ with $i=j$,
\begin{eqnarray}
\partial_{x^1}x^1&=&1+q^2x^1\partial_{x^1}+(q^2-1)(x^2\partial_{x^2}+
x^3\partial_{x^3}+x^4\partial_{x^4})+\Big(1-\frac{1}{q^3}\Big)(q^2-1)
x^5\partial_{x^5}\nonumber\\
&& \\
\partial_{x^2}x^2&=&1+q^2x^2\partial_{x^2}+(q^2-1)(x^3\partial_{x^3}+
x^5\partial_{x^5})+\Big(1-\frac{1}{q}\Big)(q^2-1)x^4\partial_{x^4} \\
\partial_{x^3}x^3&=&1+qx^3\partial_{x^3}+(q^2-1)(x^4\partial_{x^4}+
x^5\partial_{x^5})\\
\partial_{x^4}x^4&=&1+q^2x^4\partial_{x^4}+(q^2-1)x^5\partial_{x^5} \\
\partial_{x^5}x^5&=&1+q^2x^5\partial_{x^5}
\end{eqnarray}
and for $i \neq j$,
\begin{eqnarray}
\partial_{x^1}x^2=q x^2\partial_{x^1}+
\Big(\frac{1}{q^2}-1\Big)x^5\partial_{x^4}&,& \ 
\partial_{x^2}x^1=q x^1\partial_{x^2}+
\Big(\frac{1}{q^2}-1\Big)x^4\partial_{x^5}\\
\partial_{x^1}x^3=qx^3\partial_{x^1}+\frac{1-q^2}{q^{3/2}}x^5\partial_{x^3}&,& 
\ 
\partial_{x^3}x^1=qx^1\partial_{x^3}+\frac{1-q^2}{q^{3/2}}x^3\partial_{x^5}\\
\partial_{x^1}x^4=qx^4\partial_{x^1}+\frac{1-q^2}{q}x^5\partial_{x^2}&,& \
\partial_{x^4}x^1=qx^1\partial_{x^4}+\frac{1-q^2}{q}x^2\partial_{x^5} \\
\partial_{x^2}x^3=qx^3\partial_{x^2}+\frac{1-q^2}{q^{1/2}}x^4\partial_{x^3} 
&,& \
\partial_{x^3}x^2=qx^2\partial_{x^3}+\frac{1-q^2}{q^{1/2}}x^3\partial_{x^4} 
\\
\partial_{x^2}x^5=q x^5\partial_{x^2} &,& \ 
\partial_{x^5}x^2=q x^2\partial_{x^5} \\
\partial_{x^3}x^4=qx^4\partial_{x^3}, \ 
\partial_{x^3}x^5=qx^5\partial_{x^3}&,& 
\partial_{x^4}x^3=qx^3\partial_{x^4}, \ 
\partial_{x^5}x^3=qx^3\partial_{x^5}, \\
\partial_{x^4}x^5=qx^5\partial_{x^4}&,& \
\partial_{x^5}x^4=qx^4\partial_{x^5} \\
\partial_{x^1}{x^5}=x^5\partial_{x^1}, \ 
\partial_{x^5}{x^1}=x^1\partial_{x^5}&,& \
\partial_{x^2}{x^4}=x^4\partial_{x^2}, \
\partial_{x^4}{x^2}=x^2\partial_{x^4}.
\end{eqnarray}
This completes the definition of the $SO(5)_{q}$-covariant calculus on 
the quantum space.

%


\end{document}